\documentclass[twocolumn,showpacs,preprintnumbers,amsmath,amssymb]{revtex4}
\usepackage{graphicx}% Include figure files
\usepackage{color}
\begin{document}

\preprint{APS/123-QED}

\title{Berry Phase in a Single Quantum Dot with Spin-Orbit Interaction }% Force line breaks with \\
\author{Huan Wang\footnote{Email: wanghuan2626@sjtu.edu.cn}, Ka-Di Zhu\footnote{Email: zhukadi@sjtu.edu.cn }}
\address{Department of Physics, Shanghai Jiao Tong University,
Shanghai 200240, People's Republic of China}

\date{\today}
\begin{abstract}
Berry phase in a single quantum dot with Rashba spin-orbit coupling
is investigated theoretically. Berry phases as functions of magnetic
field strength, dot size, spin-orbit coupling and photon-spin
coupling constants are evaluated. It is shown that the Berry phase
will alter dramatically from $0$ to $2\pi$ as the magnetic field
strength increases. The threshold of magnetic field depends on the
dot size and the spin-orbit coupling constant.
\end{abstract}

\maketitle

\section{\label{sec:level1}Introduction}
%\lowercase{via} \textbackslash\textbackslash}
Due to its important role in encoding information, the phase of
wavefunction attracts a lot of interest in information science.
Those properties can also be used in future quantum information and
quantum computer. Thus selecting one kind of phases which can be
manipulated by quantum effect is very important. Berry phase is
believed to be a promising candidate. As a quantum mechanical system
evolves cyclically in time such that it return to its initial
physical state, its wavefunction can acquire a geometric phase
factor in addition to the familiar dynamic phase\cite{Segao,
Anandan}. If the cyclic change of the system is adiabatic, this
additional factor is known as Berry's phase\cite{Berry}, and is, in
contrast to dynamic phase, independent of energy and time.
Fuentes-Guridi et al.\cite{Fuentes} calculated the Berry phase of a
particle in a magnetic field in consideration of the quantum nature
of the light field. Yi et al.\cite{Yi} studied the Berry phase in a
composite system and showed how the Berry phases depend on the
coupling between the two subsystems. In a recent paper, San-Jose et
al.\cite{San} have described the effect of geometric phases induced
by either classical or quantum electric fields acting on single
electron spins in quantum dots. Wang and Zhu \cite{Wang H.} have
investigated the voltage-controlled Berry phases in two vertically
coupled InGaAs/GaAs quantum dots.  Most recently, observations of
Berry phases in solid state materials are
reported\cite{Yuanbo,Vartiainen,Leek}. Leek et al.\cite{Leek}
demonstrated the controlled Berry phase in a superconducting qubit
which manipulates the qubit geometrically using microwave radiation
and observes the phase in an interference experiment.

Spin-related effects have potential applications in semiconductor
devices and in quantum computation. Rashba et al.\cite{Rashba} have
described the orbital mechanisms of electron-spin manipulation by an
electric field. Sonin\cite{Sonin} has demonstrated that an
equilibrium spin current in a 2D electron gas with Rashba spin-orbit
interaction can result in a mechanical torque on a substrate near an
edge of the medium. Serebrennikov\cite{Serebrennikov} considered
that the  coherent transport properties of a charge carrier. The
transportation will cause a spin precession in zero magnetic fields
 and can be described in purely geometric terms as a
consequence of the corresponding holonomy.

 The spin-orbit interaction in
semiconductor heterstructures  is increasingly coming to be seen as
a tool which can manipulate electronic spin states
\cite{Egues,Efros}. Two basic mechanisms of the spin-orbit coupling
of 2D electrons are directly related to the symmetry properties of
QDs. They stem from the structure inversion asymmetry mechanism
described by the Rashba term\cite{Rashba,Bychkov} and the bulk
inversion asymmetry mechanism described by the Dresselhaus
term\cite{Dresselhaus}. Recently, Debald and Emary \cite{Debald}
have investigated a spin-orbit driven Rabi oscillation in a single
quantum dot with Rashba spin-orbit coupling. However, the influence
of spin-orbit interaction on Berry phase in a single quantum dot is
still lacking. In the present paper we will give a detail study on
the Berry phase evolution of  a single quantum dot with spin-orbit
interaction in a time-dependent quantized electromagnetic
environment. We will borrow quantum optics method to investigate the
impact of the spin-orbit interaction and spin-photon interaction on
Berry phase.

The paper is organized as follows. In Sec.II,  we give the model
Hamiltonian including both spin-orbit interaction and spin-photon
interaction and calculated Berry phases as functions of magnetic
field strength, dot size, spin-orbit coupling and photon-spin
coupling constants. In Sec.III, we draw the figures of the Berry
phase as a function of magnetic field strength and  some discussions
are given. The final conclusion is presented in Sec.IV.

\section{Theory}

\label{model} We consider a simple two-dimensional quantum dot
with parabolic lateral confinement potential in a perpendicular
magnetic field $\bf{B}$ which points along $z$ direction.  Then
the electron system can be described by the Hamiltonian
\cite{Debald},
\begin{eqnarray}
{H_s}=\frac{(\textbf{p}+\frac{e}{c}\textbf{A})^2}{2m^*}+\frac{m^*}{2}\omega^{2}_{0}(x^2+y^2)+\frac{1}{2}g\mu_BB\sigma_z,
\end{eqnarray}
where $\textbf{p}$ is the linear momentum operator of the
electron, $\textbf{A}(\textbf{r})=\frac{\textbf{B}}{2}(-y,x,0)$ is
the vector potential in the symmetric gauge, $\omega_0$  is the
characteristic confinement frequency, and
$\sigma=(\sigma_x,\sigma_y,\sigma_z)$ is the vector Pauli
matrices. $m^*$ is the effective mass of the electron and $g$ its
gyromagnetic factor. $\mu_{B}$ is the Bohr magneton. In the second
quantized notation, Eq.(1) becomes
\begin{eqnarray}
{H_s}=(a_x^+a_x+a_y^+a_y+1)\hbar\widetilde{\omega}+\frac{\hbar\omega_c}{2i}(a_x^+a_y-a_xa_y^+)\nonumber\\
+\frac{1}{2}g\mu_BB\sigma_z,
\end{eqnarray}
where $\omega_c=\frac{eB}{m^*c}$ and
$\widetilde{\omega}^2=\omega_0^2+\frac{\omega_c^2}{4}$. If we set
\begin{eqnarray}
a_+=\frac{1}{\sqrt{2}}(a_x-ia_y),
a_-=\frac{1}{\sqrt{2}}(a_x+ia_y),
\end{eqnarray}
Then, the Hamiltonian (2) can be written as
\begin{eqnarray}
{H_s}=n_+\hbar\omega_++n_-\hbar\omega_-+\frac{1}{2}g\mu_B
B\sigma_z,
\end{eqnarray}
where $\omega_\pm = \widetilde{\omega} \pm \omega_c/2$,
$n_+=a_+^+a_+$ and $n_-=a_-^+a_-$. In what follows we include the
spin-orbit interaction which is described as Rashba Hamiltonian in
this system \cite{Rashba}
\begin{eqnarray}
{H_{so}}=-\frac{\alpha}{\hbar}[(\textbf{p}+\frac{e}{c}\textbf
{A})\times\sigma]_{\it z},
\end{eqnarray}
where $\alpha$ is the spin-orbit coupling constant which can be
controlled by gate voltage in experiment.  On substituting Eq.(3)
into Eq.(5) and then
\begin{eqnarray}
{H_{so}}=\frac{\alpha}{\widetilde{l}}[\gamma_+(\sigma_+a_++\sigma_-a_+^+)-\gamma_-(\sigma_-a_-+\sigma_+a_-^+)],
\end{eqnarray}
where $\gamma_\pm=1\pm\frac{1}{2}(\widetilde{l}/l_B)^2$,
$\widetilde{l}=({\hbar}/{m^*\widetilde{\omega}})^{\frac{1}{2}}$
and $l_B=({\hbar}/{m^*\omega_c})^\frac{1}{2}$.

The Hamiltonians of photons and the coupling to the electron spin
can be written as follows:
\begin{eqnarray}
H_{p}=\hbar\omega_{p}b^+b,
\\ H_{p-s}=g_c(\sigma_+ +\sigma_-)(b^\dag + b),
\end{eqnarray}
where $b^+$ ($b$) and $\omega_{p}$ are the creation (annihilation)
operator and energy of the photons, respectively. $g_c$ is the
spin-photon coupling constant. Hence we obtain the total
Hamiltonian of the electron and photons:
\begin{eqnarray}
{H}=H_s+H_{so}+H_{p}+H_{p-s}\nonumber\\=\hbar\omega_+a_+^+a_++\hbar\omega_-a_-^+a_-+\frac{1}{2}g\mu_B
B\sigma_z
\nonumber\\+\frac{\alpha}{\widetilde{l}}[\gamma_+(\sigma_+a_++\sigma_-a_+^+)-\gamma_-(\sigma_-a_-+\sigma_+a_-^+)]
\nonumber\\+\hbar\omega_{p}b^+b+ g_c(\sigma_+ +\sigma_-)(b^\dag +
b).
\end{eqnarray}
Performing a unitary rotation of the spin such that
$\sigma_z\rightarrow-\sigma_z$ and
$\sigma_{\pm}\rightarrow-\sigma_{\mp}$, we arrive at the
Hamiltonian
\begin{eqnarray}
{H}=\hbar\omega_+a_+^+a_++\hbar\omega_-a_-^+a_--\frac{1}{2}g\mu_B
B\sigma_z
\nonumber\\+\frac{\alpha}{\widetilde{l}}[\gamma_-(\sigma_+a_-+\sigma_-a_-^+)-\gamma_+(\sigma_-a_++\sigma_+a_+^+)]
\nonumber\\+\hbar\omega_{p}b^+b - g_c(\sigma_+ +\sigma_-)(b^\dag +
b).
\end{eqnarray}
We now derive an approximation form of this Hamiltonian by
borrowing the observation from quantum optics that the terms
preceded by $\gamma_+$ in Eq.(10) are counterrotating, and thus
negligible under the rotating-wave approximation when the
spin-orbit coupling is small compared to the confinement
\cite{Debald}. The last term in Eq.(10) treats in the conventional
rotaing-wave approximation of quantum optics.
\begin{eqnarray}
{H}=\hbar\omega_+a_+^+a_++\hbar\omega_-a_-^+a_-+\frac{1}{2}|g|\mu_B
B\sigma_z \nonumber\\+\lambda(\sigma_+a_-+\sigma_-a_-^+)
+\hbar\omega_{p}b^+b-g_c(\sigma_+b+\sigma_-b^+),
\end{eqnarray}
where $\lambda={\alpha\gamma_-}/{\widetilde{l}}$. Since $g$ is
negative in InGaAs, we choose the absolute value $|g|$ of $g$. It
is obvious that the $\omega_{+}$ mode is decoupled from the rest
of the system, giving $H=\hbar\omega_{+}n_{+} + H_{JC}$ where
\begin{eqnarray}
H_{JC}=\hbar\omega_-a_-^+a_-+\frac{1}{2}|g|\mu_B
B\sigma_z+\hbar\omega_{p}b^+b
\nonumber\\+\lambda(\sigma_+a_-+\sigma_-a_-^+)-
g_c(\sigma_+b+\sigma_-b^+).
\end{eqnarray}
This is the well-known two mode Jaynes-Cummings model of quantum
optics. In general this Hamiltonian can not be solved exactly
except $\omega_p=\omega_-$. In what follows, for the sake of
analytical simplicity, we consider $\omega_p=\omega_-$ which we
can use a frequency-controllable laser and a special circuit to
satisfy this condition in real experiments.

 In
order to solve the above Hamiltonian, we define the normal-mode
operators:
\begin{equation}
A= e_1 a_- + e_2 b,
\end{equation}
\begin{equation}
K= e_2 a_- - e_1 b,
\end{equation}
where
\begin{equation}
e_1= \frac{\lambda}{\sqrt{\lambda^2 + g_{c}^2}}, e_2=
\frac{-g_c}{\sqrt{\lambda^2 + g_{c}^2}},
\end{equation}
with $e_1 ^2 +e_2 ^2 =1$. The new operators satisfy the commutation
relations\cite{Marchiolli}
\begin{equation}
\begin{split}
[A, A^\dag]=1, [N_A, A]=-A, [N_A, A^\dag]=A^\dag&,
\\ [K, K^\dag]=1, [N_K, K]=-K, [N_K, K^\dag]=K^\dag&,
\\ [A, K]=0, [A, K^\dag]=0, [N_A, N_K]=0&,
\end{split}
\end{equation}
where $N_A =A^\dag A(N_K =K^\dag K)$ is the number operator related
to the normal-mode operator $A (K)$. Introducing the number-sum
operator $S=N_A +N_K$ and the number-difference $D=N_A -N_K$, we can
verify that the Hamiltonian (12) transforms into the following
Hamiltonian:(i)$S=n_a +n_b$ is a conserved quantity ($n_a
=a_{-}^\dag a_{-}$ and $n_b =b^\dag b$); (ii) the operator D can be
written in terms of the generators$\{Q_+, Q_-, Q\}$ of the SU(2) Lie
algebra,
\begin{equation}
D=2(e_1^2 -e_2^2)Q_0 +2e_1 e_2 (Q_+ +Q_-),
\end{equation}
where $Q_- =a_{-}b^\dag, Q_+ =a_{-}^\dag b$, and $Q_0
=\frac{1}{2}(a_{-}^\dag a_{-} -b^\dag b)$, with $[Q_-, Q_+]=-2Q_0$
and $[Q_0, Q_\pm]=\pm Q_\pm$;(iii) the commutation relation between
the operators S and D is null, i.e., $[S, D]=0$; and consequently,
(iv) the Hamiltonian $H_{JC}$ simplifies to $H_{JC}= H_0 + V$, where
\begin{equation}
\begin{split}
 H_0 =\hbar\omega_{p}(S+\frac{1}{2}\sigma_z)&,
\\ V=\frac{1}{2}\delta\sigma_z +\lambda_A(\sigma_- A^+
+\sigma_+A)&,
\end{split}
\end{equation}
with $[H_0, V]=0$. $\lambda_A=\sqrt{\lambda^2 +g_{c}^2}$ is an
effective coupling constant and
$\delta=\omega_{p}-|g|\mu_{B}B/\hbar$. The above Hamiltonian can be
solved exactly. The eigenstates of this Hamiltonian are given by
\begin{equation}
|\Psi^{(n,\pm)} \rangle =cos\theta^{(n,\pm)}|n,\uparrow\rangle +
sin\theta^{(n,\pm)}|n+1,\downarrow\rangle,
\end{equation}
\begin{equation}
tan\theta^{(n,\pm)}=(\delta\pm\Delta_n)/2\lambda_{A}\sqrt{(n+1)},
\end{equation}
where  $\Delta_{n}=\sqrt{\delta^2+4\lambda_{A}^2(n+1)}$ and
$|\uparrow>$ ($|\downarrow>$) is the spin-up (down) state.

 According to
Ref.\cite{Fuentes}, since only the quasi-mode $A$ is coupled with
the spin of the electron,  so the phase shift operator
$U(\varphi)=e^{-i\varphi A^\dag A}$ is introduced. Applied
adiabatically to the Hamiltonian (18), the phase shift operator
alters the state of the field and gives rise to the following
eigenstates:
\begin{equation}
\begin{split}
|\psi^{(n,\pm)}>= e^{-in\varphi} cos\theta^{(n,\pm)}|n,\uparrow>+&\\
e^{-i(n+1)\varphi} sin\theta^{(n,\pm)}|n+1,\downarrow>&.
\end{split}
\end{equation}
Changing $\varphi$ slowly from 0 to $2\pi$, the Berry phase is
calculated as $\Gamma_l=i \int_0^{2\pi}
{^l}\langle\psi|\frac{\partial}{\partial\varphi}|\psi\rangle^ld\varphi$
which is given by
\begin{equation}
\Gamma_l = 2\pi[sin\theta^{(n,l)}]^2.
\end{equation}
This Berry phase is composed of two parts. One is induced by
spin-orbit interaction, the other is induced by quantized light.
Therefore if we can measure the total Berry phase and either part of
two Berry phase, we will measure the other part of Berry phase.

\section{Numerical Results }

\begin{figure}[h] %\centering
\includegraphics[width=8cm]{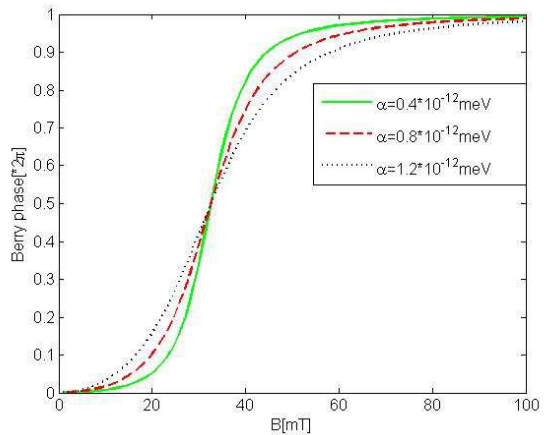}
\caption{The Berry phase $\Gamma_+$ as a function of magnetic field
strength $B$ with three spin-orbit coupling constants (
$\alpha=0.4\times10^{-12}eVm$, $0.8\times10^{-12}eVm$ and
$1.2\times10^{-12}eVm$). The other parameters used are $g=-4$,
$m^*/m_e =0.05$, $g_c=0.01meV$, $l_0 =80nm$, and $n=0$.
}\label{fig:sect1:learning3}
\end{figure}

\begin{figure}[h] %\centering
\includegraphics[width=8cm]{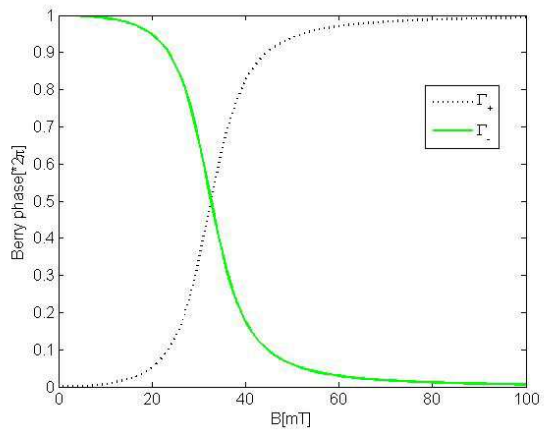}
\caption{The Berry phases of $\Gamma_+$ and $\Gamma_-$ as a function
of magnetic field strength $B$. The parameters used are
$\alpha=0.4\times10^{-12}eVm$, $g=-4$, $m^*/m_e =0.05$,
$g_c=0.01meV$, $l_0 =80 nm$, and $n=0$. }\label{fig:sect1:learning3}
\end{figure}

For the illustration of the numerical results, we choose the typical
parameters of the InGaAs: $g=-4$, $m^*/m_e =0.05$ ($m_e$ is the mass
of free electron). The dot size is defined by
$l_{0}=\sqrt{\hbar/m^{*}\omega_{0}}$. Figure 1 depicts the Berry
phases $\Gamma_{+}$ as a function of the magnetic field strength for
three spin-orbit couplings. In Figure 1, we can find that all the
Berry phases change almost from 0 to $2\pi$ as the magnetic field
strength varies from $20 mT$ to $50 mT$.  When other parameters are
fixed, the spin-orbit coupling constant changes as
$\alpha=0.4\times10^{-12}eVm$, $0.8\times10^{-12}eVm$ and
$1.2\times10^{-12}eVm$, the Berry phases $\Gamma_+$ will have a
slight movement in the figure. When $B<20mT$ and $B>50mT$, the Berry
phase changes gradually, while when $20 mT<B<50 mT$, the Berry phase
changes dramatically. As the coupling constant increases, the Berry
phase changes from sharply to slowly. The Sh\"{o}rdinger equation
has two different eigenenergies when $n=0$. The two eigenenergies
will give two different Berry phases. Figure 2 illustrates these two
Berry phases. In Figure 2, when the others parameter are fixed, one
of the Berry phase changes from 0 to $2\pi$ , while the other
changes from $2\pi$ to 0 as the magnetic field strength varies from
$20 mT$ to $50 mT$. Two Berry phases have an intersecting point at
approximatively $B=33 mT$, which is corresponding to the resonant
point.

\begin{figure}[h] %\centering
\includegraphics[width=8cm]{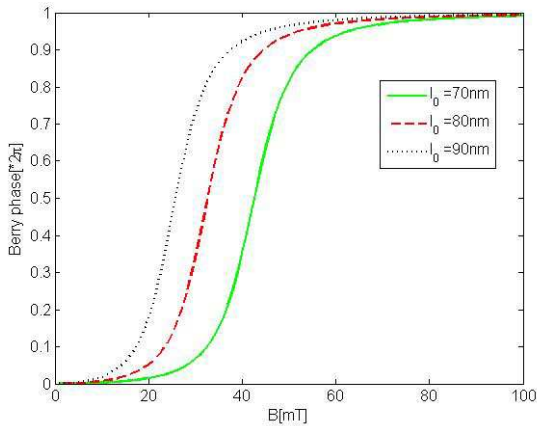}
\caption{The Berry phases $\Gamma_+$ as a function of $B$ with
three three different dot sizes ($l_0 =70nm, 80nm, 90nm$). The
parameters used are $\alpha=0.4\times10^{-12}eVm$, $g=-4$, $m/m_e
=0.05$, $g_c=0.01meV$, and $n=0$. }\label{fig:sect1:learning3}
\end{figure}

\begin{figure}[h] %\centering
\includegraphics[width=8cm]{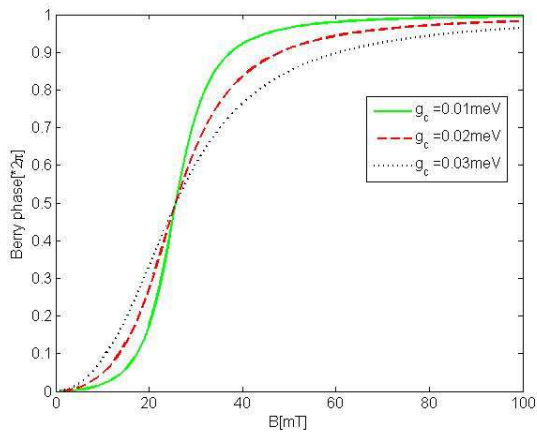}
\caption{The Berry phases $\Gamma_+$ as a function of $B$ with
three different light coupling constants ($g_c=0.01meV, 0.02meV,
0.03meV$). The parameters used are $\alpha=0.4\times10^{-12}eVm$,
$g=-4$, $m/m_e =0.05$, $l_0 =80 nm$, and $n=0$.
}\label{fig:sect1:learning3}
\end{figure}
Figure 3 shows the effect of the dot size on the Berry phase. When
dot size becomes large from 70nm to 90nm, although all three Berry
phases change from 0 to $2\pi$, the threshold points of the magnetic
field have a large movement. When the dot size is 70nm, the Berry
phase will change dramatically at approximately 40mT, while the dot
sizes are 80nm and 90nm, the turning points are approximately at
30mT and 20mT, respectively.  This implies that the bigger the dot,
the smaller the threshold of the magnetic field strength. Figure 4
illustrates the influence of spin-photon coupling constant on Berry
phase.  As the coupling constant becomes large, the Berry phase
becomes less drastic as shown in Figure 4.

In a recent paper, Giuliano et al.\cite{Giuliano} have designed an
experimental arrangement, which is capacitively coupled the dot to
one arm of a double-path electron interferometer. The phase carried
by the transported electrons may be influenced by the dot. The dot's
phase gives raise to an interference term in the total conductance
across the ring. More recently, Leek et al.\cite{Leek} have measured
Berry phase in a Ramsey fringe interference experiment.  Our
experimental setup proposed here is analogous with these two
arrangements as shown in Figure 5.  A beam light is split into two
beams, one of the beams passes through the dot, and interferes with
the other one. Accurate control of the light field for dot is
achieved through phase and amplitude modulation of laser radiation
coupled to the dot. We choose a special designed electric circuit to
ensure the magnetic and laser vary synchronistically. Through
detecting the interfered light, we can measure the Berry phase.

\begin{figure}[h] %\centering
\includegraphics[width=8cm]{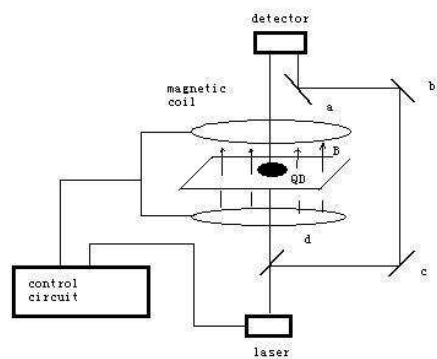}
\caption{A sketch of a possible experimental setup to detect the
Berry phase.  a, b and c are three mirror, d is a beam splitter.
}\label{fig:sect1:learning3}
\end{figure}

\section{Conclusions}
In conclusion, we have theoretically investigated the Berry phase in
a single quantum dot in the presence of Rashba spin-orbit
interaction. Berry phases as functions of magnetic field strength,
dot size, spin-orbit coupling and photon-spin coupling constants are
evaluated. It is shown that for a given quantum dot, the spin-orbit
coupling constant and photon-spin coupling constant the Berry phase
will alter dramatically from $0$ to $2\pi$ as the magnetic field
strength increases. The threshold of magnetic field is dependent on
the Rashba spin-orbit coupling constant, spin-photon coupling
constant and the dot size. We also propose a practicable method to
detect the Berry phase in such a quantum dot system. Finally, we
hope that our predictions in the present work can be testified by
experiments in the near future.

\begin{acknowledgments}
This work has been supported in part by National Natural Science
Foundation of China (No.10774101) and the National Ministry of
Education Program for Training PhD.
\end{acknowledgments}

\end{document}